\documentclass[10pt, conference, letterpaper]{IEEEtran}
\usepackage{cite}
\usepackage{amsmath,amssymb,amsfonts}
\usepackage{graphicx}
\usepackage{textcomp}
\usepackage{xcolor}
\usepackage[T1]{fontenc}

\usepackage{blindtext}
\usepackage[numbers]{natbib}
\usepackage{url}
\usepackage{cleveref}
\usepackage{floatrow}
\usepackage[tableposition=top]{caption}
\usepackage{float}
\usepackage{booktabs}
\floatstyle{plaintop}
\restylefloat{table}

\usepackage[normalem]{ulem}

\usepackage{algorithm}
\usepackage{algpseudocode}
\usepackage{algorithmicx}
\algdef{SE}[DOWHILE]{Do}{doWhile}{\algorithmicdo}[1]{\algorithmicwhile\ #1}%

\algnewcommand\algorithmicforeach{\textbf{for each}}
\algdef{S}[FOR]{ForEach}[1]{\algorithmicforeach\ #1\ \algorithmicdo}

\usepackage{subcaption}

\usepackage{tikz}
\usepackage{pgfplots}
\usepackage{pgf}
\pgfplotsset{width=7.75cm,compat=1.9}
\usepackage[utf8]{inputenc} 

\usepackage{tablefootnote}
\usepackage{threeparttable}
\usepackage{multirow}
\usepackage{booktabs}
\usepackage{harveyballs}

\usepackage{siunitx}

\usepackage{minibox}
\usepackage{tcolorbox}

\usepackage{setspace}
\def\BibTeX{{\rm B\kern-.05em{\sc i\kern-.025em b}\kern-.08em
    T\kern-.1667em\lower.7ex\hbox{E}\kern-.125emX}}
\newlength{\bibitemsep}
\newlength{\bibparskip}
\let\oldthebibliography\thebibliography
\renewcommand\thebibliography[1]{%
  \oldthebibliography{#1}%
  \setlength{\parskip}{\bibitemsep}%
  \setlength{\itemsep}{\bibparskip}%
}

\newcommand{\algoname}{Algorithm}
\newcommand{\linename}{line}

\usepackage{cleveref} 
\crefname{section}{§}{§§}
\Crefname{section}{§}{§§}

\definecolor{tumColorRed}{HTML}{D95117} 

\definecolor{tumColorRedPlot}{HTML}{ea7237} 
\definecolor{tumColorGreen}{HTML}{9fba36} 
\definecolor{tumColorBlue}{HTML}{5e94d4} 
\definecolor{tumColorPink}{HTML}{9B468D} 

\definecolor{colorTBV}{HTML}{fcad03}
\definecolor{colorTBA}{HTML}{24cc06}
\definecolor{tumColorLightBlue}{HTML}{f0f5fa}   

\usepackage{xcolor}

\long\def\ignore#1{}

\RequirePackage{caption}   

\renewcommand{\baselinestretch}{0.987}

\AtBeginDocument{
        \captionsetup{font={footnotesize}}     

        \setlength{\textfloatsep}{1.3\baselineskip plus  0.2\baselineskip minus  0.4\baselineskip}    
        \setlength\belowcaptionskip{0pt}                

        \setlength\abovecaptionskip{5cm}     
}

\usepackage[nolist]{acronym}

\begin{acronym}

\acro{iot}[IoT]{Internet of Things}
\acro{mitm}[MitM]{Machine-in-the-Middle}
\acro{dos}[DoS]{Denial of Service}
\acro{tsn}[TSN]{Time Sensitive Networking}
\acro{ptp}[PTP]{Precision Time Protocol}
\acro{gnss}[GNSS]{Global Navigation Satellite System}
\acro{bmca}[BMCA]{Best Master Clock Algorithm}
\acro{oc}[OC]{Ordinary Clock}
\acro{tc}[TC]{Transparent Clock}
\acro{bc}[BC]{Boundary Clock}
\acro{sdn}[SDN]{Software-defined Networking}
\acro{rtt}[RTT]{Round-Trip Time}
\acro{nic}[NIC]{Network Interface Card}
\acro{dpdk}[DPDK]{Data Plane Development Kit}
\acro{pps}[PPS]{Pulse-Per-Second}

\end{acronym}

\begin{document}

\title{PTPsec: Securing the Precision Time Protocol Against Time Delay Attacks Using Cyclic Path Asymmetry Analysis}

\author{
\IEEEauthorblockN{Andreas Finkenzeller\IEEEauthorrefmark{1}, Oliver Butowski\IEEEauthorrefmark{1}, Emanuel Regnath\IEEEauthorrefmark{2}, Mohammad Hamad\IEEEauthorrefmark{1}, and Sebastian Steinhorst\IEEEauthorrefmark{1}}
\IEEEauthorblockA{
\IEEEauthorrefmark{1}Technical University of Munich, Germany\\
\IEEEauthorrefmark{2}Siemens AG, Germany\\
Email: \IEEEauthorrefmark{1}firstname.lastname@tum.de, \IEEEauthorrefmark{2}firstname.lastname@siemens.com
}}

\maketitle

\begin{abstract}

High-precision time synchronization is a vital prerequisite for many modern applications and technologies, including Smart Grids, Time-Sensitive Networking (TSN), and 5G networks.
Although the Precision Time Protocol (PTP) can accomplish this requirement in trusted environments, it becomes unreliable in the presence of specific cyber attacks.
Mainly, time delay attacks pose the highest threat to the protocol, enabling attackers to diverge targeted clocks undetected.
With the increasing danger of cyber attacks, especially against critical infrastructure, there is a great demand for effective countermeasures to secure both time synchronization and the applications that depend on it.
However, current solutions are not sufficiently capable of mitigating sophisticated delay attacks.
For example, they lack proper integration into the PTP protocol, scalability, or sound evaluation with the required microsecond-level accuracy.
This work proposes an approach to detect and counteract delay attacks against PTP based on cyclic path asymmetry measurements over redundant paths.
For that, we provide a method to find redundant paths in arbitrary networks and show how this redundancy can be exploited to reveal and mitigate undesirable asymmetries on the synchronization path that cause the malicious clock divergence.
Furthermore, we propose PTPsec, a secure PTP protocol and its implementation based on the latest \mbox{IEEE 1588-2019} standard.
With PTPsec, we advance the conventional PTP to support reliable delay attack detection and mitigation.
We validate our approach on a hardware testbed, which includes an attacker capable of performing static and incremental delay attacks at a microsecond precision. Our experimental results show that all attack scenarios can be reliably detected and mitigated with minimal detection time.

\end{abstract}


\section{Introduction}

Precise time synchronization is indispensable for many applications and technologies, such as Smart Grids, \ac{tsn}, and 5G networks.
To work correctly, these applications usually require synchronization accuracies on a microsecond or sub-microsecond level, and even small deviations can already have significant impacts. 
The consequences range from performance degradation to, in the worst case, full system failure. 
For example, delay attacks in Smart Grids render the control loops unstable, eventually disrupting the entire system \cite{sargolzaei2014delayed}.

The \ac{ptp} (see \cref{sec:ptp}) is a commonly used network protocol to provide accurate time synchronization in the aforementioned scenarios. 
Unfortunately, \ac{ptp} was initially designed without any security considerations in mind and, hence, makes certain network assumptions, such as path symmetry and message integrity, that might not always hold. Especially in spatially large networks like Smart Grids, for instance, cyber attacks targeting the network infrastructure are becoming more likely, forcing the protocol designers to reconsider security concepts in \ac{ptp}.
With the four-pronged approach in Annex P of the latest revision from 2019 \cite{ieee-1588-2019}, the \ac{ptp} standard aims to address this problem. While the suggested cryptographic countermeasures can guarantee message integrity and protect against many other attack vectors, including message replay and spoofing attacks, they cannot counteract all known threats.
In particular, time delay attacks (see \cref{sec:time_delay_attack}) remain an unsolved problem so far, as previous work has extensively shown \cite{barreto2016, annessi2018, finkenzeller2022feasible}.

\begin{figure}[t]
    \centering
    \includegraphics[width=\linewidth]{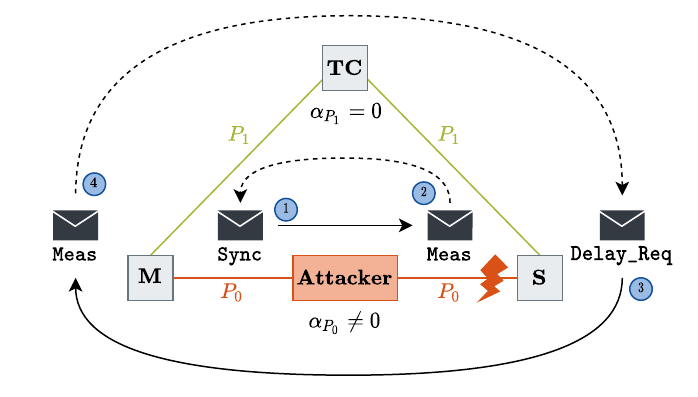}
    \caption{The path asymmetry analysis concept to detect time delay attacks within our proposed PTPsec protocol.
    The PTP \texttt{Sync} message, sent via the attacked path $P_0$ (1), is followed by our newly introduced \texttt{Meas} message over the genuine path $P_1$ to complete the first round trip (2). Similarly, the exchange of the \texttt{Delay\_Req} (3) and another \texttt{Meas} message (4) leads to a second circulation. This allows for cyclic RTT measurements from which we derive the current path asymmetry $\alpha_{P_0}$ to reveal and mitigate ongoing delay attacks.}
    \label{fig:system_overview}
\end{figure}

Delay attacks exploit and violate the protocol's assumption of symmetric path delays by maliciously introducing
unidirectional delays into the network.
In the literature, initial solutions to counteract delay attacks against PTP have been proposed recently, none of which provide an effective solution to the problem yet.
Threshold-based approaches, for example, as presented in \cite{yang2013time, li2021security}, perform the attack detection by means of continuously monitoring the reported path delay and comparing it to a previously defined threshold.
However, the threshold definition is quite challenging. Static thresholds must be set very conservatively to avoid false alarms, while dynamic thresholds, which adapt based on previous measurements, cannot adequately protect the system from incremental delay attacks.
Moreover, the reported path delay is not guaranteed to change at all during a successful delay attack \cite{neyer2019redundant}.
Other solutions, including additional network guards \cite{moussa2015detection, alghamdi2017advanced} or special monitoring and reporting mechanisms \cite{moussa2020extension, moradi2021new} only work within a limited attacker model.
Because the detection method is not sufficiently coupled to the time synchronization, attackers can distinguish \ac{ptp} messages from other network traffic and handle them differently to bypass the existing mitigations.
Besides, there are first efforts to countermeasures based on path redundancy.
However, the available works \cite{mizrahi2012game, neyer2019redundant} only present some general ideas that, among other things, lack scalability and proper protocol integration to become applicable.
Hence, there is still a great need for a secure and comprehensive solution that entirely protects \ac{ptp} from time delay attacks.

\textit{Contributions:}
This work investigates cyclic \ac{rtt} measurements to analyze network path asymmetries and the applicability for delay attack detection and mitigation in time synchronization protocols.
The cyclic analysis is enabled by dedicated measurement packets that we introduce in our proposed \ac{ptp}sec protocol to entangle the attack detection with the synchronization procedure.
Each \ac{ptp} event message invokes a subsequent measurement packet that is returned to the originator via a redundant path to complement the cyclic \ac{rtt} measurement as depicted in \figurename~\ref{fig:system_overview}.
To finally mitigate ongoing delay attacks, we analyze the path asymmetry based on the two \ac{rtt} measurements obtained in one \ac{ptp} synchronization round and compensate for it accordingly.
To the best of our knowledge, \ac{ptp}sec is the first protocol that efficiently detects and mitigates time delay attacks.
In particular, we:
\begin{itemize}
    \item analyze cyclic \ac{rtt} measurements and show their effectiveness for reliable path asymmetry identification (\cref{sec:asymmetry_measurements}),
    \item derive a theoretical model for delay attack detection in arbitrary networks (\cref{sec:attack_detection}),
    \item present \ac{ptp}sec, a secure \ac{ptp} protocol and its implementation as an extension of the latest IEEE 1588-2019 standard that adopts our proposed attack mitigation method (\cref{sec:secure_ptp}), and
    \item validate the \ac{ptp}sec implementation on a realistic hardware testbed to confirm that our approach successfully detects and mitigates time delay attacks (\cref{sec:evaluation}). 
\end{itemize}

\section{System Model}

\begin{figure}
    \centering
    \includegraphics[width=0.95\linewidth]{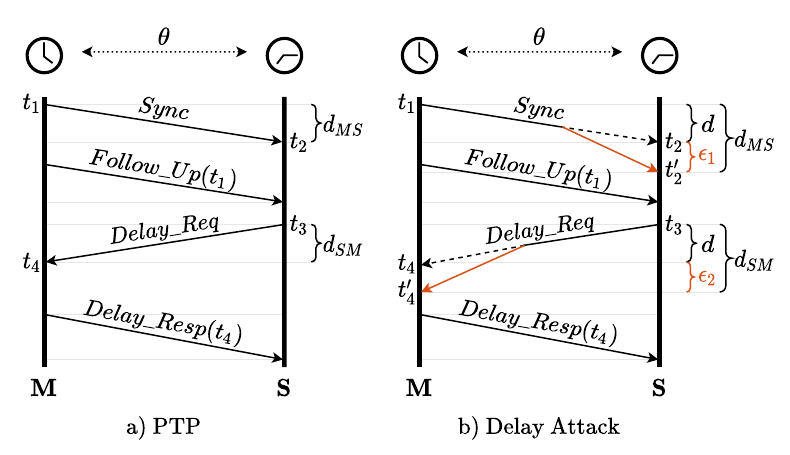}
    \caption{a) PTP message flow with timestamping to minimize the clock offset $\theta$. b) When attackers can delay PTP event messages (\texttt{Sync} or \texttt{Delay\_Req}), they create path asymmetries that impair the clock synchronization.}
    \label{fig:ptp_protocol}
\end{figure}

\subsection{Precision Time Protocol}
\label{sec:ptp}

The precision time protocol, as defined in the IEEE 1588 standard \cite{ieee-1588-2019}, is a network protocol that synchronizes devices with an accuracy of less than \SI{1}{\micro\second}. 
Thereby, the standard considers various clock types to account for different device behaviors in common \ac{ptp} networks.
Particularly, \acp{oc} denote master and slave nodes\footnote{In this work, we stick to the terminology used in the official standard to avoid any confusion besides being potentially inappropriate.} which either provide a time reference to others or synchronize their clocks to a specific master, respectively.
In contrast, \acp{tc} do not actively synchronize to other clocks but are transparent switching devices that connect multiple \acp{oc} to form a large network.
The synchronization procedure comprises two phases.
In the protocol's first phase, all participating nodes announce their existence and determine the best (i.e., most accurate) clock in the network, denoted the \textit{Grandmaster} clock, using the \ac{bmca}.
In the second phase, all other nodes start actively synchronizing their clocks to the \textit{Grandmaster} by periodically exchanging \ac{ptp} messages.
Along with the exchange of \texttt{Sync} and \texttt{Delay\_Req} messages, four timestamps are captured at both the transmission and reception of each packet. Here, the required use of hardware timestamping increases the synchronization accuracy significantly.
The two so-called event messages are critical for the achieved synchronization accuracy since the obtained timestamps are the basis for subsequent offset calculations.
With the additional assumption of symmetric path delays ($d_{MS} = d_{SM}$), which is an indispensable assumption for \ac{ptp}, the current clock offset $\theta$ can be computed as:

\begin{equation}
\label{eq:ptp_offset_formula}
    \theta = \frac{(t_2 - t_1) - (t_4 - t_3)}{2}
\end{equation}

\noindent where $t_1$, $t_2$, $t_3$, and $t_4$ denote the captured timestamps of the two event messages. The diagram in \figurename~\ref{fig:ptp_protocol}a depicts the captured timestamps along the entire message flow of one synchronization round.
Note that the \texttt{Follow\_Up} message conveying the captured transmission timestamp $t_1$ is only required when \ac{ptp} operates in two-step mode.

\subsection{Time Delay Attack}
\label{sec:time_delay_attack}

However, if the synchronization path is not symmetric because of malicious network activities, for example, the offset calculations are based on an invalid assumption, leading to erroneous synchronization results.
So, if attackers were able to deliberately delay the \texttt{Sync} and \texttt{Delay\_Req} messages by $\epsilon_1$ and $\epsilon_2$ respectively with $\epsilon_1 \neq \epsilon_2$, as illustrated in Fig.~\ref{fig:ptp_protocol}b, the clock offset $\theta'$ would mistakenly compute as:

\begin{align}
\label{eq:ptp_offset_attack_formula}
\begin{split}
        \theta' &= \frac{((t_2 \color{tumColorRed} + \epsilon_1 \color{black}) - t_1) - ((t_4 \color{tumColorRed} + \epsilon_2 \color{black}) - t_3)}{2} \\ 
            &= \frac{(t_2 - t_1) - (t_4 - t_3)}{2} \color{tumColorRed} + \frac{\epsilon_1 - \epsilon_2}{2} \\
            &= \theta \color{tumColorRed} + \frac{\alpha}{2}
\end{split}
\end{align}

\noindent with $\alpha = d_{MS} - d_{SM} = \epsilon_{1} - \epsilon_{2}$.
Provided $\alpha = 0$, the path is perfectly symmetric, and the time synchronization is successful. 
However, if $\alpha \neq 0$, the path becomes asymmetric, and the synchronized clock runs behind or ahead in time of the \textit{Grandmaster} clock, respectively.
The described time delay attack can be implemented in multiple ways.
For example, Barreto et al. \cite{barreto2016} present a static delay attack scenario, where the attackers deliberately extend the optical fibers in smart grid networks to add a constant one-way delay, turning the synchronization path asymmetric.
In \cite{finkenzeller2022feasible}, the authors perform an incremental delay attack, where the introduced asymmetric path delay gradually increases, to showcase more dynamic attacks.
Both works show the general feasibility of time delay attacks and further emphasize their devastating impact on \ac{ptp}, highlighting the great need for sound countermeasures.

\subsection{Network Model}
\label{sec:graph_modeling}

A typical \ac{ptp} deployment comprises an entire network, including many devices that shall be synchronized.
Therefore, we model a given \ac{ptp} network as the graph $\mathcal{G}~=~(V, E)$ with $V = \{v_0, \ldots , v_n\}, \; n \in \mathbb{N}$ denoting the set of vertices and $E = \{e_0, \ldots , e_m\} \subseteq V \times V, \; m \in \mathbb{N}$ the set of edges.
Each vertex $v_i \in V$ represents a node participating in the \ac{ptp} protocol. 
For proper synchronization, we require at least the elected \textit{Grandmaster} node and one additional slave device that synchronizes to the master. Thus, we assume our vertex set $V$ to contain these two nodes at the minimum, which we denote for further reference as $M \in V$ for the master and $S \in V$ for the slave node, respectively.
All edges $e_i \in E$ are considered bidirectional and model the connecting links between all involved network devices.
Furthermore, we assume $\mathcal{G}$ to be connected. Hence, at least one path connects $M$ and $S$ for the \ac{ptp} message exchange, which we denote as $P_0$.
Also, we extend the notation of $\alpha$ to express the asymmetry $\alpha_{e_i}$ of a specific edge $e_i \in E$. Additionally, we define the path asymmetry $\alpha_{P_i}$, which is experienced by all traversing packets on path $P_i$, as the sum of the individual link asymmetries of all composing links:
\begin{equation}
\label{eq:path_asymmetry_definition}
    \alpha_{P_i} = \sum_{j} \alpha_{e_j}, \,\forall e_j \in P_i.
\end{equation}

Finally, we assume full control of the packet routing, which can be achieved, for example, by leveraging \ac{sdn} capabilities \cite{li2021security}.

\subsection{Attacker Model}
\label{sec:attacker_model}

There exist many attacks against time synchronization protocols, particularly against \ac{ptp} \cite{alghamdi2021}. In this work, however, we mainly focus on delay attacks and related attacker capabilities due to their criticality.
Similar to \cite{barreto2016, finkenzeller2022feasible}, we deny the attackers internal access to the protocol, i.e., the possibility to directly modify \ac{ptp} header fields.
If this was possible, the attackers were powerful enough to manipulate the synchronization on a protocol level by changing timestamps and the correction field contents, which would render delay attacks superfluous. Security protocols, such as IPsec and MACsec, can be appropriate remedies to ensure message integrity.
Furthermore, attackers are not allowed to compromise hosts or intermediary network devices in a way that would grant them internal protocol access or allow them to update the clocks directly with a similar reasoning.

Nevertheless, we assume the attackers to have full knowledge of the network topology, including knowledge about deployed \ac{ptp} clock types. Additionally, they are fully aware of any implemented protection methods (e.g., IPsec).
After compromising a link, the attackers can delay all passing packets individually.
More precisely, they can deliberately add unidirectional delays to selected messages in both transmission directions independently.
This fine-grained packet delay is even possible with specific protection mechanisms enabled, such as traffic encryption, as shown in \cite{annessi2018}.
Moreover, the attackers are able to simultaneously perform multiple delay attacks precisely synchronized at different locations in the network if they compromise more than one link. 
There are no restrictions on the number of compromised links or the strategy of how attackers perform joint attacks. 
However, we assume that the selection of hijacked links will remain constant for a sufficiently long period to allow for steady-state analysis.
Particularly, the attackers may freely change the introduced delay for an attacked link over time but not attack a different link.
Finally, we require all nodes that participate in the \ac{ptp} protocol to be honest, i.e., to follow the protocol as specified in the standard or in our proposed PTPsec protocol, respectively.

\section{Asymmetry Analysis}
\label{sec:asymmetry_measurements}

If we could precisely measure unidirectional delays in the network, time delay attacks would be easily detected.
However, accurate one-way path delay measurements are a challenging problem in network theory \cite{choi2005one}.
In the following analysis, we benefit from the work of Gurewitz et al. \cite{gurewitz2001estimating, gurewitz2006one} on one-way delay estimation which we extend to derive our cyclic path asymmetry analysis. 
Additionally, we propose a general path asymmetry model that forms the basis for the delay attack detection method presented in \cref{sec:attack_detection}.
First, we start with a simple two-node example from which we develop the general model for arbitrary networks.

\subsection{Two-Node Scenario}
\label{sec:two_node_scenario}

Given a simple synchronization scenario with only two nodes $M$ and $S$ and one attacked edge $e_0$, as shown in \figurename~\ref{fig:delay_meas_approach}a, we are trying to estimate the link asymmetry $\alpha_{e_0} \neq 0$.
However, it is impossible to reliably determine $\alpha_{e_0}$ with only one available link.
Thus, we need at least one additional link to obtain a circular structure that we can use for further analysis.

\subsubsection{One redundant link}

\begin{figure}
    \centering
    \includegraphics[width=0.9\textwidth]{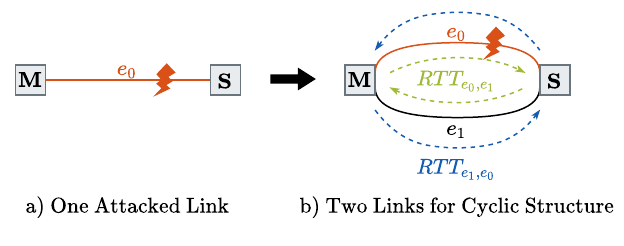}
    \caption{Cyclic path asymmetry analysis illustrated with two nodes. While there is no efficient method to calculate the asymmetry with only one link (a), a second link enables a cyclic structure for further analysis (b). Two RTT measurements in opposing directions can be smartly combined to determine the link asymmetry $\alpha_{e_0}$ of the attacked link $e_0$.}
    \label{fig:delay_meas_approach}
\end{figure}

If there exists a second edge $e_1$ which is redundant to $e_0$, both edges form a cycle, as depicted in \figurename~\ref{fig:delay_meas_approach}b.
Now, we can perform a cyclic \ac{rtt} measurement using $e_0$ and $e_1$ by forwarding a packet from $M$ to $S$ on link $e_0$ which is subsequently returned to $M$ via $e_1$. Upon arrival at node $M$, the elapsed time $RTT_{e_0,e_1}$ computes as:

\begin{equation}
    RTT_{e_0,e_1} = t_{in} - t_{eg}
\end{equation}

\noindent where $t_{in}$ is the packet's measured ingress and $t_{eg}$ the egress timestamp at node $M$.
Additionally, we conduct a second \ac{rtt} measurement $RTT_{e_1,e_0}$ with reverse transmission direction, i.e., forwarding the packet on link $e_1$ to $S$ and returning it via $e_0$.
Both measurements include unidirectional delays of $e_0$, however, in opposite directions. 
For $RTT_{e_0,e_1}$, edge $e_0$ contributes in forward direction ($M$ to $S$) while $RTT_{e_1,e_0}$ contains its delay in backward direction ($S$ to $M$).
If we further assume that link $e_1$ is perfectly symmetric ($\alpha_{e_1} = 0$), i.e., the contributed link delay is equal in both directions, we can calculate the asymmetry of link $e_0$ as: 

\begin{equation}
\label{eq:asymmetry_calculation}
    \alpha_{e_0} = RTT_{e_0,e_1} - RTT_{e_1,e_0}
\end{equation}

\noindent Note that the opposing one-way delays of link $e_1$ cancel out due to its symmetry and we are left with the desired difference of both unidirectional delays of $e_0$.
Here, the essential part is the cyclic measurement approach capturing the opposing one-way delays of a single link isolated in distinct measurements.
Furthermore, the two \acp{rtt} measurements originate from the same clock avoiding the need for any time synchronization.
Also note that the same measurements could have likewise been taken on node $S$ since the starting point in a cyclic measurement is irrelevant.

\subsubsection{N redundant links}
In the previous example with one redundant edge, the asymmetry measurement only worked because we assumed the additional link $e_1$ to be symmetric.
However, if the second link was also asymmetric due to another delay attack, for example, attackers could thwart the measurement attempt while still successfully introducing unidirectional delays on $e_0$.
For that, they need to add similar one-way delays $\epsilon$ to both links $e_0$ and $e_1$ but in opposing directions, so that the delays cancel out each other in \eqref{eq:asymmetry_calculation} while the individual links become asymmetric.
In such a case, we need another edge $e_2$ with symmetric link delay for compensation. This additional link increases the number of cycles in our network allowing for further independent cyclic measurements.
With two redundant edges, we can perform two asymmetry measurements that include $e_0$ leading to:

\begin{align}
\begin{split}
    \alpha^{(1)} = RTT_{e_0,e_1} - RTT_{e_1,e_0} \\
    \alpha^{(2)} = RTT_{e_0,e_2} - RTT_{e_2,e_0}
\end{split}
\end{align}

\noindent with $\alpha^{(1)}$ and $\alpha^{(2)}$ being independent estimates for the targeted link asymmetry $\alpha_{e_0}$.
Although $\alpha^{(1)}$ might equal to zero indicating perfect link symmetry due to the mentioned attack strategy, $\alpha^{(2)}$ will yield the correct estimate because of the demanded link symmetry of $e_2$ and the reasoning of \eqref{eq:asymmetry_calculation}.

Similarly, we can derive the general case of $n$ redundant edges (in addition to $e_0$) for the two-node scenario. 
From $2n$ pairwise \ac{rtt} measurements $RTT_{e_0,e_i}$ and $RTT_{e_i,e_0}$,\, $i \in [1, n]$, we get $n$ estimates for the link asymmetry $\alpha_{e_0}$:

\begin{align}
\begin{split}
    \alpha^{(1)} &= RTT_{e_0,e_1} - RTT_{e_1,e_0} \\
    \alpha^{(2)} &= RTT_{e_0,e_2} - RTT_{e_2,e_0} \\
    &\hspace{0.5cm}\vdots \\
    \alpha^{(n)} &= RTT_{e_0,e_n} - RTT_{e_n,e_0}
\end{split}
\end{align}

\noindent Furthermore, we can state the following important finding:
\begin{tcolorbox}[colback=tumColorLightBlue, colframe=black, boxrule=.2mm, boxsep=0.5mm, sharp corners=all]
    \textbf{Finding 1:} In order to successfully determine $\alpha_{e_0}$, we need at least one link $e_i,\, i \in [1,n]$ with symmetric link delay $\alpha_{e_i} = 0$, which is redundant to $e_0$. 
    Using this symmetric link $e_i$, the asymmetry measurement yields a correct estimate 
    $\alpha^{(i)} = RTT_{e_0,e_i} - RTT_{e_i,e_0} = \alpha_{e_0}$.
\end{tcolorbox}

\noindent This necessary condition directly results from the previous reasoning that attackers could introduce various unidirectional delays to cancel out each other in the cyclic measurements if we allowed all links to be asymmetric.

\subsection{Multi-Node Scenario}

\begin{figure}
    \centering
    \includegraphics[width=0.9\textwidth]{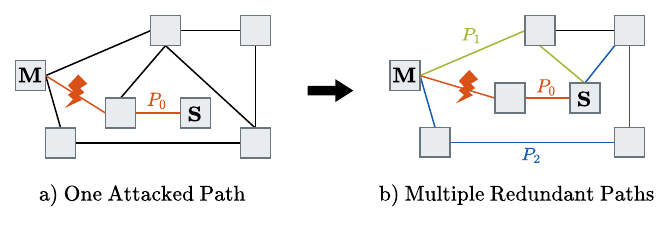}
    \caption{Redundant path principle in multi-node networks. To efficiently estimate the path asymmetry $\alpha_{P_0}$, we require other edge-disjoint paths $P_i$ to enable cyclic \ac{rtt} measurements.}
    \label{fig:delay_meas_approach_general}
\end{figure}

Realistic networks usually comprise more than two nodes. Therefore, we need to derive a general path asymmetry model that is applicable to networks of any size.
Similar to the two-node scenario, we attempt to estimate the asymmetry $\alpha_{P_0}$ on the synchronization path $P_0$. 
We equally require at least one redundant path $P_1$ in addition to $P_0$, as exemplified in \figurename~\ref{fig:delay_meas_approach_general}, to enable the cyclic measurements previously proposed in \cref{sec:two_node_scenario}.
Note that $P_1$ must not share any edge with $P_0$ to be considered fully redundant since all edges of $P_0$ are contributing to its total path asymmetry as defined in \eqref{eq:path_asymmetry_definition}. If a single joint edge existed, it would affect both paths equally and make a cyclic measurement with independent forward and backward delays impossible.
Therefore, we need $P_0$ and $P_1$ to be edge-disjoint: $P_0 \cap P_1 = \emptyset$.
%
\noindent With this requirement, we can formulate the general approach to estimate path asymmetries in arbitrary networks.
Given the network $\mathcal{G} = (E, V)$, the synchronization path $P_0$, and $n$ redundant paths $P_{i},\, i \in [1,n]$ between nodes $M$ and $S$, we perform $2n$ cyclic \ac{rtt} measurements for each pair $P_{0}$ and $P_{i},\, i \in [1, n]$. 
From these measurements, we derive the following set of equations to get $n$ estimates for the path asymmetry $\alpha_{P_0}$:
\begin{align}
\label{eq:general_path_asymmetry_measurement}
\begin{split}
    \alpha^{(1)} &= RTT_{P_{0},P_{1}} - RTT_{P_{1},P_{0}} \\
    \alpha^{(2)} &= RTT_{P_{0},P_{2}} - RTT_{P_{2},P_{0}} \\
    & \hspace{0.5cm} \vdots \\
    \alpha^{(n)} &= RTT_{P_{0},P_{n}} - RTT_{P_{n},P_{0}} \\
\end{split}
\end{align}

\noindent Moreover, we state the generalized finding from our path asymmetry analysis in arbitrary networks:
\begin{tcolorbox}[colback=tumColorLightBlue, colframe=black, boxrule=.2mm, boxsep=0.5mm, sharp corners=all]
    \textbf{Finding 2:} For all symmetric paths $P_i,\, i \in [1,n]$ with $\alpha_{P_i} = 0$, we get correct estimates $\alpha^{(i)} = RTT_{P_{0},P_{i}} - RTT_{P_{i},P_{0}} = \alpha_{P_0}$.
    Hence, we require at least one redundant path to be symmetric in order to successfully determine the desired path asymmetry $\alpha_{P_0}$ for the synchronization path $P_0$.
\end{tcolorbox}

\section{Attack Detection and Mitigation}
\label{sec:attack_detection}

\subsection{Detection Criteria}

From \eqref{eq:ptp_offset_attack_formula}, we know that successful time delay attacks result in non-zero path asymmetries. Hence, we can now use the previously derived path asymmetry model to detect and mitigate these attacks in \ac{ptp} networks.
With $n+1$ total edge-disjoint paths $P_i,\, i \in [0,n]$ connecting nodes $M$ and $S$, where $P_0$ is used for \ac{ptp} synchronization, we yield $n$ independent asymmetry estimates for $\alpha_{P_0}$, as stated in \eqref{eq:general_path_asymmetry_measurement}. 
Based on these estimates, we define the detection criterion for an ongoing delay attack on path $P_0$ impeding the time synchronization between $M$ and $S$ as follows:

\vspace{-0.3cm}

\begin{align}
    \begin{split}
        \textit{delay\_attack} \;\leftarrow\; 
        \begin{cases}
            True, & \exists \, \alpha^{(i)} \neq 0, \, i \in [1, n] \\
            False,  & \text{otherwise}
        \end{cases}
    \end{split}
\end{align}

\noindent Additionally, we know that the cyclic measurements only work if at least one path is symmetric. 
Thus, we can successfully detect delay attacks that include up to $n$ attacked paths.
Furthermore, we can try to determine the actual position of the attacked paths by analyzing the estimated path asymmetries $\alpha^{(i)}$ and search for attack configurations that match the obtained result. 
It turns out that we can cluster all paths into two groups, where one group is either entirely genuine and the other fully attacked or vice versa.
However, this decision cannot be made without further assumptions, such as limiting the number of attackers to an upper bound of $\#\,attackers \leq \lfloor \frac{n}{2} \rfloor$.

\subsection{Attack Mitigation}

If an attack is detected, we can furthermore use the measured path asymmetry $\alpha_{P_0}$ of the synchronization path $P_0$ to mitigate the ongoing attack. For that, we calculate the rectified clock offset $\theta_{rect}$ that is compensating malicious path asymmetries using the reported \ac{ptp} offset $\theta_{rep}$ from \eqref{eq:ptp_offset_formula} and the insight gained from \eqref{eq:ptp_offset_attack_formula}:

\begin{equation}
\label{eq:rectified_clock_offset}
    \theta_{rect} = \theta_{rep} - \frac{\alpha_{P_0}}{2}
\end{equation}

\noindent This rectified clock offset can be used as input for \ac{ptp}'s control algorithm to securely update the local clock oscillator despite any ongoing attack.

\subsection{Finding Redundant Paths}
\label{sec:redundant_paths_finding}

\begin{algorithm}[t!]
    \caption{Adapted Ford-Fulkerson Algorithm}
    \label{alg:ford_fulkerson}
    \begin{algorithmic}[1]
        \Require Network $\mathcal{G} = (V, E)$, Source $M \in V$, Sink $S \in V$
        \Ensure Set of edge-disjoint paths $\mathcal{P}$ from $M$ to $S$
        \State $paths \gets \emptyset$  \label{alg:ford_fulkerson:start_init}
        \State $flow(e) \gets 0 \enspace \textbf{for each } e \in E$ \label{alg:ford_fulkerson:end_init}
        \While {$p \gets \text{find } M{\text -}S \text{ path with } flow(e) = 0 \enspace \forall e \in p$} \label{alg:ford_fulkerson:path_search}
            \State $paths.\text{insert(}p\text{)}$ \label{alg:insert path}
            \State $flow(e) \gets 1 \enspace \textbf{for each } e \in p$ \label{alg:update flow}
        \EndWhile
        \State \Return $paths$ \label{alg:ford_fulkerson:return}
    \end{algorithmic}
\end{algorithm}

Another essential aspect of the presented method is finding redundant paths in a given network $\mathcal{G}$ which results to finding edge-disjoint paths to obtain full redundancy.
For that, we present an approach to derive all eligible paths by leveraging both Menger's theorem \cite{bohme2001menger} and the Max-Flow-Min-Cut theorem \cite{ford1956maximal}.
Menger states there are $k$ pairwise edge-disjoint $M{\text -}S$ paths if and only if $S$ is still reachable from $M$ after removing $k-1$ arbitrary edges from the graph \cite{bohme2001menger}.
Hence, we are interested in the minimum edge cut $k$ that is required to disconnect the two nodes $M$ and $S$ in $\mathcal{G}$.
By introducing the non-negative capacity function $c(e) = 1$, which assigns each edge $e_i \in E$ the maximum capacity of one, this problem is equivalent to maximizing the flow from $M$ to $S$, as stated by the Max-Flow-Min-Cut theorem. 
Finally, the maximum flow can be efficiently computed by the Ford-Fulkerson algorithm \cite{ford2015flows} in polynomial time. 
Thus, to derive the number of edge-disjoint $M{\text -}S$ paths in a given \ac{ptp} network $\mathcal{G}$, we propose an adapted version of the Ford-Fulkerson algorithm which is depicted in \algoname~\ref{alg:ford_fulkerson}.
First, we initialize the set of existing paths and the assigned flow values for each edge to zero (\linename s~\ref{alg:ford_fulkerson:start_init}-\ref{alg:ford_fulkerson:end_init}).
Then, we repeatedly search for $M\text{-}S$ paths containing only edges with remaining capacity, i.e., with $flow(e) = 0$ (\linename~\ref{alg:ford_fulkerson:path_search}) and insert them into the set of existing paths (line~\ref{alg:insert path}). The search can be accomplished, for example, with the breadth-first search algorithm.
For each found path, we additionally update the flow values of all comprising edges to mark their capacity as consumed (\linename~\ref{alg:update flow}) for subsequent search iterations.
Finally, we return the resulting set of desired edge-disjoint paths (\linename~\ref{alg:ford_fulkerson:return}).

\subsection{Multipoint Synchronization}

In typical \ac{ptp} deployments, many nodes synchronize to a single \textit{Grandmaster} clock. Thus, we need to find redundant paths for all eligible synchronization paths. 
For this purpose, we run \algoname~\ref{alg:ford_fulkerson} for every node, which shall be synchronized to the \textit{Grandmaster} before the normal protocol flow. Moreover, we can leverage \ac{sdn} capabilities to dynamically recalculate redundant paths on any network change.

\section{Proposed PTPsec}
\label{sec:secure_ptp}

\begin{figure}[t]
    \centering
    \includegraphics[width=0.9\linewidth]{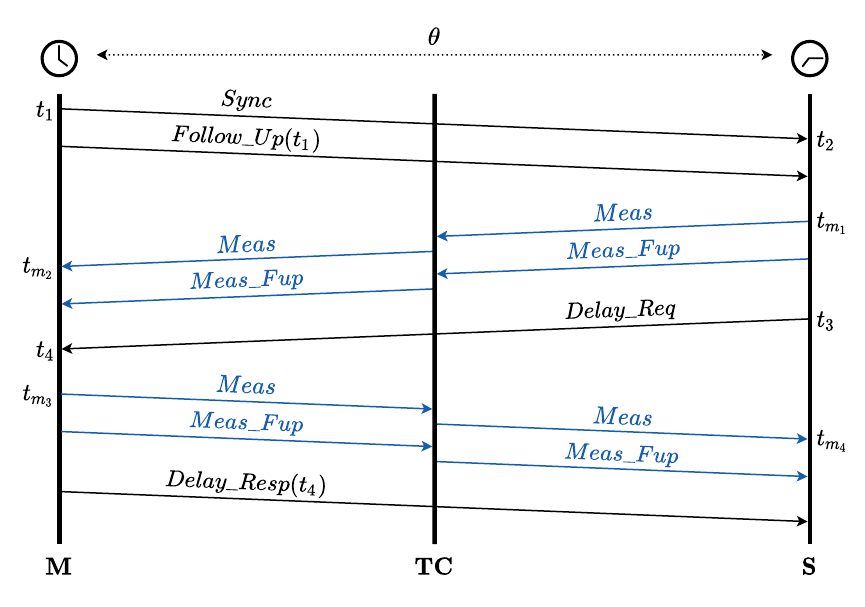}
    \caption{Proposed PTPsec message flow to protect against time delay attacks. After the reception of PTP event messages (\texttt{Sync} and \texttt{Delay\_Req}), dedicated (\texttt{Meas}) messages are returned to the originator via a redundant network path, here indicated with the additional TC, to enable cyclic path asymmetry measurements. The \texttt{Meas\_Fup} messages convey the captured timestamps $t_{m_1}$ and $t_{m_3}$, respectively, when PTP operates in two-step mode.}
    \label{fig:secure_ptp_protocol}
\end{figure}

As already highlighted in \cref{sec:ptp}, the latest IEEE 1588-2019 standard (\ac{ptp} v2.1) is not secure against time delay attacks.
Hence, we propose the \ac{ptp}sec protocol, which integrates appropriate countermeasures into the existing IEEE standard to improve \ac{ptp}'s resilience against such attacks.
Recall that in our attacker model, attackers can precisely delay individual packets. Particularly, they could selectively delay only \ac{ptp} messages if these were distinguishable and independent from our measurement packets.
Therefore, we need to entangle the \ac{rtt} measurements with the \ac{ptp} synchronization procedure to prevent bypassing the presented attack detection and mitigation approach.
For that, we aim to integrate the two critical event messages \texttt{Sync} and \texttt{Delay\_Req} directly into the \ac{rtt} measurements.
As a consequence, we ensure that deliberate packet delays, which impede the time synchronization, also equally affect the proposed detection approach.

Let $RTT_{P_{0},P_{i}}$ be the \ac{rtt} measurement including the synchronization path $P_{0}$ and a redundant path $P_i$ to estimate the path asymmetry $\alpha_{P_0}$. 
Then, the \texttt{Sync} message, which is sent from node $M$ to $S$ via $P_0$, already initiates the first \ac{rtt} measurement. 
Once received at node $S$, an immediate response is returned to $M$ via the redundant path $P_i$ to finish the round trip.
Since this response has a different purpose than existing \ac{ptp} message types, we introduce the new message type \texttt{Meas} in our proposed PTPsec protocol.
Similar to \texttt{Sync} messages, also \texttt{Meas} packets are \ac{ptp} event messages due to the requested timestamps at packet transmission and reception. We denote the captured \texttt{Meas} timestamps with an additional $m$, e.g., $t_{m_1}$, to distinguish them from the four default \ac{ptp} timestamps.
If the protocol operates in two-step mode, we require an additional new message \texttt{Meas\_Fup} to forward the captured timestamp in a separate follow-up message.
To continue the message flow, $S$ sends the expected \texttt{Delay\_Req} message via $P_0$ to $M$ which is immediately followed by another \texttt{Meas} message to $S$ via $P_i$ upon reception to finish the second measurement cycle yielding  $RTT_{P_{i},P_{0}}$.
Finally, $M$ answers the request with a \texttt{Delay\_Resp} message to complete the synchronization protocol.
\figurename~\ref{fig:secure_ptp_protocol} shows the full sequence of our proposed \ac{ptp}sec protocol.
With all captured timestamps in this iteration, the two \ac{rtt} measurements compute as:

\begin{align}
\begin{split}
    RTT_{P_{0},P_{i}} &= (t_{m_2} - t_1) - (t_{m_1} - t_2) \\
    RTT_{P_{i},P_{0}} &= (t_{m_4} - t_3) - (t_{m_3} - t_4) 
\end{split}
\end{align}

\noindent from which we derive the path asymmetry estimate for $P_0$:

\begin{equation}
    \alpha_{P_0} = RTT_{P_{0},P_{i}} - RTT_{P_{i},P_{0}} 
\end{equation}

\noindent Note that in the presented procedure, both critical \ac{ptp} event messages are fully integrated into the path asymmetry analysis. Consequently, any malicious delay equally impacts the synchronization and the detection mechanism, preventing attackers from bypassing our approach even if individual \ac{ptp} messages can be distinguished and delayed.
Since the proposed \ac{ptp}sec protocol requires additional measurement packets for the asymmetry analysis, it also introduces a message overhead compared to conventional \ac{ptp}. In particular, four supplementary packets are sent on each of the $n$ redundant network path per \ac{ptp} synchronization cycle, resulting in the following total number of \ac{ptp}sec messages per cycle:

\begin{equation}
\label{eq:packet_overhead}
    \#\,packets = 4 + 4n
\end{equation}

\noindent Note the packet count's linear increase with the number of redundant paths.
However, the actual number of redundant paths to consider for securing the protocol is a security parameter that can be adjusted depending on the assumed attack model.
Furthermore, \ac{ptp} messages usually account only for a small part of the entire network traffic which alleviates the impact of the packet increase.

\section{Evaluation}
\label{sec:evaluation}

\begin{figure}[t!]
    \centering
    \includegraphics[width=0.9
    \textwidth]{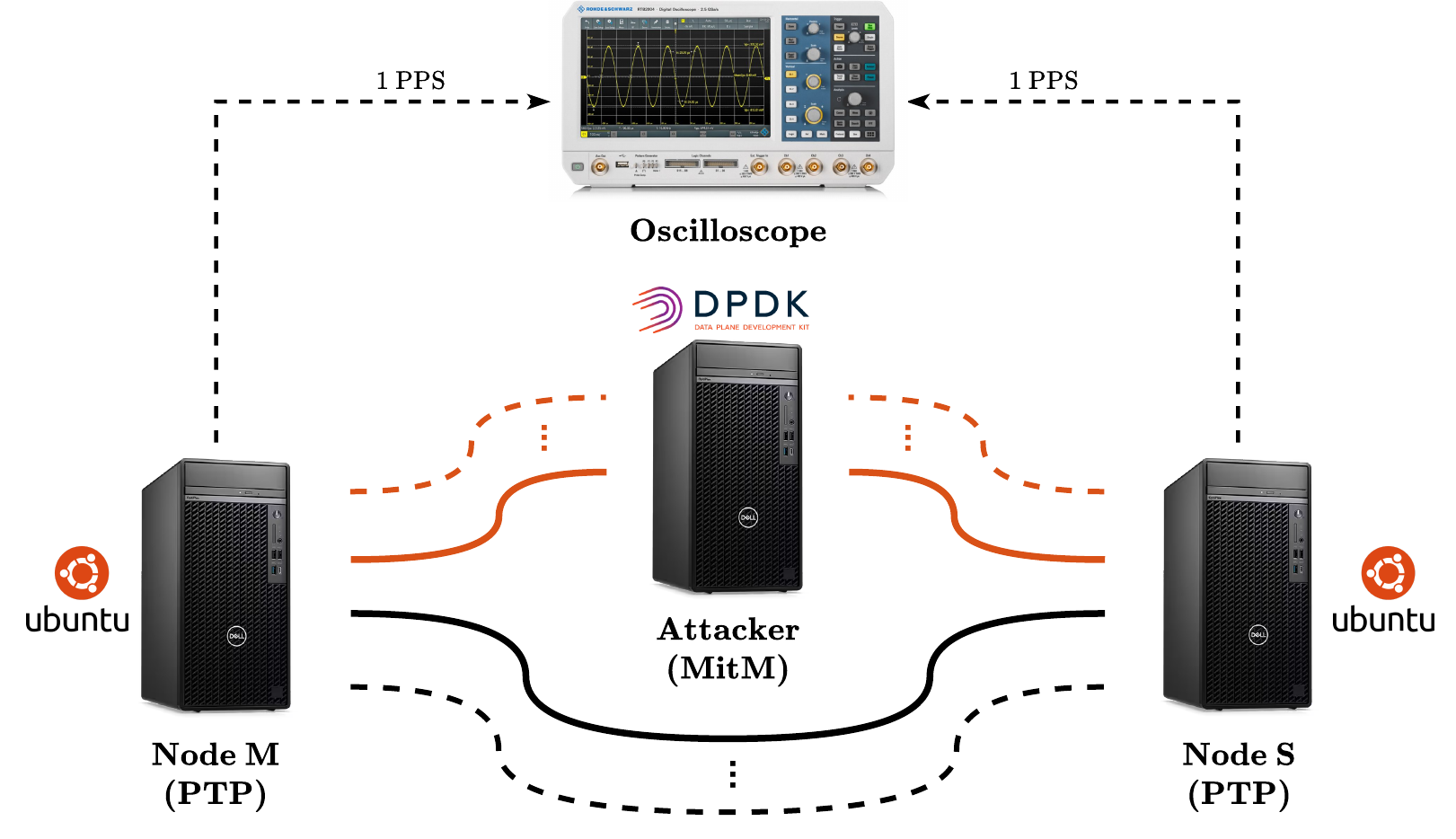}
    \caption{Hardware testbed for comprehensive evaluation of our proposed approach under realistic circumstances. The synchronization path is interrupted by a MitM attacker delaying selected PTP event messages in transit to diverge the slave clock. We closely monitor the malicious clock offset with the oscilloscope by comparing the generated PPS signals.}
    \label{fig:test_setup}
\end{figure}

For reliable evaluation, we validate the performance of our proposed \ac{ptp}sec protocol under various realistic attack scenarios on our hardware testbed.

\subsection{Hardware Setup}

The hardware setup consists of two PCs serving as master and slave to be synchronized. Both machines are operating on \textit{Ubuntu 20.04} and run an implementation of our \ac{ptp}sec protocol, which is based on \textit{linuxptp-v3.1.1}.
The two \acp{oc} use Intel i210 \acp{nic} for IEEE 1588-compliant hardware timestamping support and are connected with wired Ethernet connections.
The synchronization path is interrupted by an attack device running a \ac{dpdk} application to act as a transparent L2 switch.
Moreover, it exploits its \ac{mitm} position to deliberately delay either \texttt{Sync} or \texttt{Delay\_Req} messages to implement an effective time delay attack against \ac{ptp}, as described in \cref{sec:time_delay_attack}.
The other path is a direct point-to-point connection and constitutes a redundant path that is considered genuine and symmetric for all following experiments.
Extending the setup with more redundant paths follows the same principle, as illustrated in \figurename~\ref{fig:test_setup}.
To monitor the actual clock offset between master and slave, we compare the generated \ac{pps} signal phases on a Rohde\&Schwarz RTB oscilloscope.

\subsection{Asymmetry Detection}

In total, we present three experiments to validate the detection performance of our protocol.
We attack the two relevant \ac{ptp} event messages \texttt{Sync} and \texttt{Follow\_Up} and assume that \texttt{Meas} and \texttt{Meas\_Fup} are only sent via the symmetric redundant path.
Attacking the newly introduced measurement packets instead of the \ac{ptp} messages would lead to similar results because for the asymmetry analysis, it is irrelevant which specific path is attacked provided one genuine path exists.

\subsubsection{Static Delay (\texttt{Sync})}

\begin{figure}[t!]

    \centering
    \scalebox{0.45}{\input{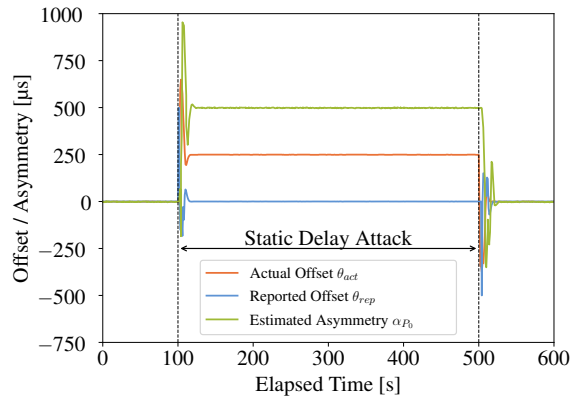}}
    \caption{Static delay attack with $\epsilon_1 = \SI{500}{\micro\second}$ targeting the PTP \texttt{Sync} message between $t = \SI{100}{\second}$ and $t = \SI{500}{\second}$. The difference between the measured clock offset (orange) and the reported clock offset (blue) confirms the ongoing attack. Nevertheless, our proposed asymmetry analysis successfully detects this malicious behavior (green).}
    \label{fig:static_delay_sync}
\end{figure}

In the first experiment, we start with a static attack scenario, where the attacker adds a constant delay of $\epsilon_1 = \SI{500}{\micro\second}$ to all passing \texttt{Sync} messages. 
The attack starts at $t = \SI{100}{\second}$ and lasts until $t = \SI{500}{\second}$, as shown in \figurename~\ref{fig:static_delay_sync}.
During that period, we observe that the actual clock offset increases to $\theta_{act} = \SI{250}{\micro\second}$, i.e., by half the introduced one-way delay, while the reported \ac{ptp} clock offset $\theta_{rep}$ remains at zero. 
This behavior clearly shows the success of the delay attack because the actual clock offset changes to the expected value given by \eqref{eq:ptp_offset_attack_formula} without being reported by conventional \ac{ptp}.
Furthermore, the estimated synchronization path asymmetry rises to $\alpha_{P_0} = \SI{500}{\micro\second}$ during the attack period. 
Since $\alpha_{P_0} > 0$, the path delay from node $M$ to node $S$ is higher than in the opposite direction, and we conclude an ongoing delay attack targeting the \texttt{Sync} message. Hence, the applied attack was successfully detected.

\subsubsection{Static Delay (\texttt{Delay\_Req})}

\begin{figure}[t!]
    \centering
    \scalebox{0.45}{\input{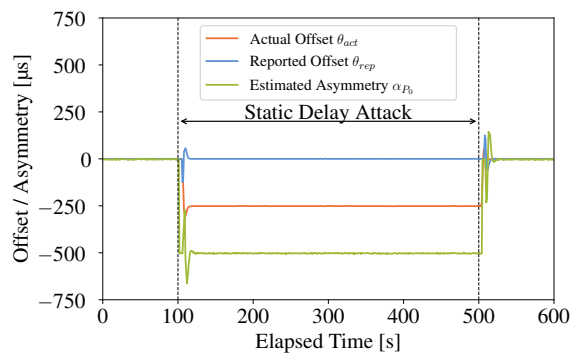}}
    \caption{Static delay attack with $\epsilon_2 = \SI{500}{\micro\second}$ targeting the PTP \texttt{Delay\_Req} message between $t = \SI{100}{\second}$ and $t = \SI{500}{\second}$. The difference between the measured clock offset (orange) and the reported clock offset (blue) confirms the ongoing attack. However, our proposed asymmetry analysis successfully detects this malicious behavior (green).}
    \label{fig:static_delay_dreq}
\end{figure}

We repeat the previous attack in the second experiment with the only difference in the targeted \ac{ptp} message. This time, the attacker delays passing \texttt{Delay\_Req} messages instead, which results in a path asymmetry in the opposite direction.
As \figurename~\ref{fig:static_delay_dreq} depicts, the reported \ac{ptp} clock offset $\theta_{rep}$ remains at zero while the actual clock offset $\theta_{act} = \SI{-250}{\micro\second}$ and the estimated path asymmetry $\alpha_{P_0} = \SI{-500}{\micro\second}$ both show a change of sign. 
The negative values match our expectations due to the different \ac{ptp} message that was targeted. 
These results confirm that the attack detection works independently of the present asymmetry direction.

\subsubsection{Incremental Delay Attack}

\begin{figure}[t!]
    \centering
    \scalebox{0.45}{\input{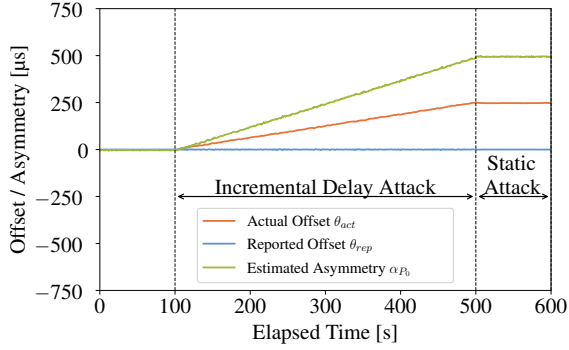}}
    \caption{Incremental delay attack, where the offset is gradually increased by $\Delta \epsilon = \SI{1.25}{\micro\second}$ starting at $t = \SI{100}{\second}$ with $\epsilon_1 = \SI{0}{\micro\second}$ to a final level of $\epsilon_1 = \SI{500}{\micro\second}$. The attack targets the PTP \texttt{Sync} message. 
    The difference between the measured clock offset (orange) and the reported clock offset (blue) confirms the ongoing attack. Nevertheless, our proposed asymmetry analysis successfully detects this malicious behavior (green).}
    \label{fig:incremental_delay_sync}
\end{figure}

Finally, we change the attack method to an incremental delay attack to evaluate the detection performance more dynamically.
Therefore, we gradually increase the packet delay for passing \texttt{Sync} messages to slowly diverge the slave's clock. 
The attack starts at $t = \SI{100}{\second}$. Each second, we increase the added delay by $\Delta\epsilon = \SI{1.25}{\micro\second}$ to eventually reach a level of $\epsilon_1 = \SI{500}{\micro\second}$ at $t = \SI{500}{\second}$ which remains as a static offset until the end of the experiment.
From the results in \figurename~\ref{fig:incremental_delay_sync}, we observe that not only the actual clock offset gradually ascends from $\theta_{act} = \SI{0}{\micro\second}$ to $\theta_{act} = \SI{250}{\micro\second}$, but also the estimated path asymmetry $\alpha_{P_0}$ slowly follows the incremental delay to reach a final level of $\alpha_{P_0} = \SI{500}{\micro\second}$.
Therefore, we conclude that our proposed protocol also reliably detects incremental delay attacks.

\subsection{Attack Mitigation}

\begin{figure}[t!]
    \centering
    \scalebox{0.45}{\input{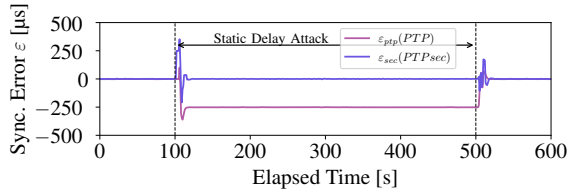}}
    \caption{Remaining clock error after delay attack compensation considering the estimated path asymmetry.}
    \label{fig:attack_compensation}
\end{figure}

As the previous experiments show, our proposed \ac{ptp}sec protocol reliably detects the malicious path asymmetry in all attack scenarios.
To further evaluate the attack mitigation performance, we analyze the synchronization errors $\varepsilon_{ptp}$ and $\varepsilon_{sec}$ of both conventional \ac{ptp} and \ac{ptp}sec, respectively, during a static delay attack by comparing the reported clock offset to the actual clock offset measured by the oscilloscope. For \ac{ptp}sec, we use the rectified offset from \eqref{eq:rectified_clock_offset}.
As the results in \figurename~\ref{fig:attack_compensation} illustrate, \ac{ptp}sec reliably mitigates the attack.

\subsection{Detection Time}

\begin{figure}[t!]
    \centering
    \scalebox{0.47}{\input{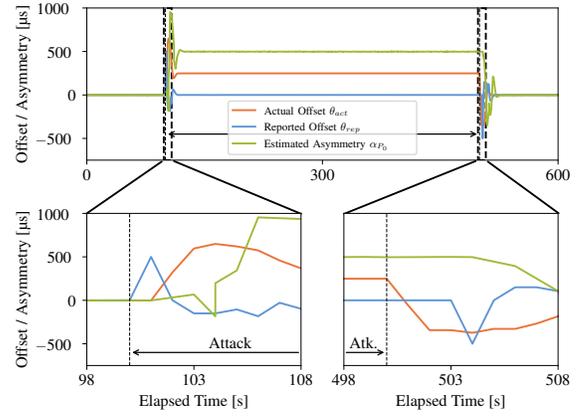}}
    \caption{Timing analysis of our attack detection approach. The magnified plots at the bottom illustrate our method's fast detection time at the start and the end of the applied attack.}
    \label{fig:response_time}
\end{figure}

In addition to the general detection and mitigation capabilities, we further examine the timing behavior of the proposed approach.
Particularly, we investigate the elapsed time between the start of the attack and its visibility in the measured path asymmetry, i.e., the detection time of our method.
This time is critical since it potentially opens a small window for attackers during which the deployed attack is effective without an appropriate system response and, thus, needs to be minimized.
In our setup, the \texttt{Sync} message timeout interval is set to \SI{1}{\second} which also predetermines the clock update and asymmetry measurement rate to approx. this period.
As the plot in \figurename~\ref{fig:response_time} shows, the measured path asymmetry follows the actual clock offset almost immediately within less than five time steps at the start and the end of the attack.

\section{Discussion}
\label{sec:discussion}

The proposed \ac{ptp}sec protocol has successfully detected and counteracted the delay attacks in all conducted experiments validating the effectiveness of our approach.
Since the chosen scenarios are representative of all known time delay attack strategies, we can conclude that our cyclic asymmetry analysis enables reliable attack mitigation provided redundant paths are available in the network.
Interestingly, this redundancy requirement is already enforced by some communication standards, such as \ac{tsn}, so our secure protocol could be deployed in related applications at no cost.
In all other cases, the introduced procedure in \algoname~\ref{alg:ford_fulkerson} can serve as a network analysis tool to evaluate the current security level regarding time delay attacks. Once a bottleneck has been identified, the network can be selectively patched to meet the specified requirements.
Regarding the protocol overhead of \ac{ptp}sec, the increased security is not for free. Since both the number of successfully detected asymmetric paths and the required packet count scale with the number of redundant network paths, it is a trade-off between desired security and acceptable network load that system designers have to make.

\section{Related Work}
\label{sec:related_work}

Attack detection and prevention in time synchronization have been broadly covered by research in recent years.
The proposed solutions to protect \ac{ptp} from time delay attacks can be mainly clustered into three groups.
First, there exist threshold-based detection techniques. The idea is to continuously monitor the reported clock offset and path delay and decide whether or not the system is under attack based on a defined threshold. The threshold can be either static, as proposed by Ullmann et al. \cite{ullmann2009delay}, or dynamically updated over time, as presented in \cite{yang2013time, li2021security}.
However, the performance of these techniques strongly depends on the chosen threshold, which is quite challenging.
The second category comprises solutions that rely on special guard devices and additional reporting systems deployed in the network.
Alghamdi et al. \cite{alghamdi2020cyber} present the idea of a trusted supervisor node that performs anomaly detection in the network to detect ongoing attacks.
Moussa et al. \cite{moussa2015detection} propose a similar idea, including a dedicated guard node that participates in the time synchronization protocol but does not update its clock. Instead, it only compares the results to another time reference to reveal potential delay attacks.
However, the guard node only secures its own synchronization path, and other nodes that do not share the same path are still vulnerable.
In \cite{moussa2020extension}, Moussa et al. refine their initial proposal and introduce a more sophisticated approach, where all nodes additionally send timestamped reports to a specific network time reference node. These reports are exchanged by means of custom event messages that are not related to the \ac{ptp} protocol stack.
Thus, sophisticated attackers can distinguish report messages from actual \ac{ptp} messages and react differently to disrupt time synchronization without being noticed.
Moradi et al. \cite{moradi2021new} present a similar method, including a dedicated reporting scheme for attack detection.
The third class of countermeasures utilizes path redundancy for time delay attack protection.
There exist some works, such as \cite{mizrahi2012slave, komes2012ieee, shpiner2016multipath}, which cover path redundancy for network-based time synchronization to improve the provided synchronization accuracy.
While not explicitly mentioning any security aspects, the works already indicate the capabilities of redundant path approaches, which could also be adopted in the security domain. 
In \cite{mizrahi2012game}, Mizrahi proposes redundant paths as countermeasure for delay attacks and presents a theoretical analysis of this idea. 
Furthermore, Neyer et al. \cite{neyer2019redundant} perform supplementary experiments on a hardware testbed to showcase the general applicability of redundant paths as viable improvement for time synchronization security.
\begin{table}[t!]
    \begin{threeparttable}
    	\caption{Delay Attack Detection Approaches Comparison}
        \label{tab:related_work_table}
        \renewcommand{\arraystretch}{1.2} 
        \setlength{\tabcolsep}{4pt} 
        \begin{tabular}{l c c c c c c}
    		\toprule
    		\multirow{2}{*}{\rotatebox[origin=c]{0}{\textbf{Solution}}} &
    		\multirow{2}{*}{\rotatebox[origin=c]{0}{\textbf{Approach}}} &
    		\multicolumn{3}{c}{\textbf{Attack Resistance}} &
    		\multirow{2}{*}{\rotatebox[origin=c]{0}{\textbf{Scal.}}} &
    		\multirow{2}{*}{\rotatebox[origin=c]{0}{\textbf{Exp.}}} \\
    		&
    		&
    		\rotatebox[origin=c]{0}{Static} &
    		\rotatebox[origin=c]{0}{Incr.}  &
    		\rotatebox[origin=c]{0}{Select.}  &
            & \\
    		\midrule
        	\cite{yang2013time} & Threshold & \harveyBallFull & \harveyBallHalf & \harveyBallFull & \harveyBallNone & \harveyBallHalf \\
        	\cite{ullmann2009delay} & Threshold & \harveyBallFull & \harveyBallHalf & \harveyBallFull & \harveyBallNone & \harveyBallNone \\
        	\cite{li2021security} & Threshold & \harveyBallFull & \harveyBallHalf & \harveyBallFull & \harveyBallHalf & \harveyBallFull \\
        	\cite{alghamdi2020cyber} & Guard & \harveyBallFull & \harveyBallFull & \harveyBallNone & \harveyBallNone & \harveyBallNone \\
        	\cite{moussa2015detection} & Guard &  \harveyBallFull & \harveyBallFull & \harveyBallFull & \harveyBallNone & \harveyBallHalf \\
        	\cite{moussa2020extension} & Guard &  \harveyBallFull & \harveyBallFull & \harveyBallNone & \harveyBallFull & \harveyBallHalf \\
        	\cite{moradi2021new} & Guard & \harveyBallFull & \harveyBallFull & \harveyBallNone & \harveyBallFull & \harveyBallHalf \\
        	\cite{mizrahi2012game} & Path Red. & \harveyBallFull & \harveyBallFull & \harveyBallFull & \harveyBallFull & \harveyBallNone \\
        	\cite{neyer2019redundant} & Path Red. & \harveyBallFull & \harveyBallFull & \harveyBallFull & \harveyBallNone &  \harveyBallFull \\
    		\textbf{PTPsec} & \textbf{Path Red.} & \harveyBallFull & \harveyBallFull & \harveyBallFull & \harveyBallFull & \harveyBallFull \\
    		\bottomrule
    		\multicolumn{5}{l}{  \harveyBallNone\ no or n/a, \harveyBallHalf\ partially, \harveyBallFull\ yes}
    	\end{tabular}
    \end{threeparttable}
\end{table}

\textit{Comparative Analysis:} 
Table~\ref{tab:related_work_table} presents an additional comparison between our proposed solution and existing works. The comparison is based on the adopted approach, the ability to detect static, incremental, and selective delay attacks, the scalability of the proposed solution, and the experimental validation method employed (hardware setup, simulation, or no validation). 
While all the solutions are capable of detecting static attacks, \cite{yang2013time, ullmann2009delay,li2021security} fall short of fully detecting incremental attacks due to the threshold-based approach. Others, like \cite{alghamdi2020cyber, moussa2020extension, moradi2021new}, are still vulnerable to selective delay attacks, where the attackers distinguish between \ac{ptp} and other network traffic.
In contrast to \ac{ptp}sec, many other solutions do not adequately scale either due to the need for a recurring setup phase after each network change, as seen in \cite{yang2013time, ullmann2009delay}, including single points of failure \cite{alghamdi2020cyber, moussa2015detection}, or due to limitations in the system model \cite{neyer2019redundant}.
Furthermore, hardware testbed validation is essential for ensuring the effectiveness of proposed detection methods, which most of the compared solutions are lacking.
Based on Table~\ref{tab:related_work_table}, the three most comparable solutions to our system are \cite{li2021security}, \cite{mizrahi2012game}, and \cite{neyer2019redundant}.
However, \cite{li2021security} cannot reliably detect incremental delay attacks because of the threshold-based approach while additionally requiring an initial learning phase that invalidates with any network change.
Furthermore, the presented solutions in \cite{mizrahi2012game} and \cite{neyer2019redundant} lack experimental validation and scalability, respectively.
Moreover, while partially providing sound detection approaches, none of the compared works can actually mitigate time delay attacks which is an exclusive feature of PTPsec.
 
\section{Conclusion}
\label{sec:conclusion}

This work introduces a theoretical model for cyclic path asymmetry analysis that can be efficiently used for time delay attack mitigation.
The derived attack detection method reliably reveals both static and incremental time delay attacks provided the existence of at least one redundant network path with symmetric delay.
With \ac{ptp}sec, we present a protocol that secures the latest IEEE 1588 standard against time delay attacks by incorporating our proposed mitigation techniques on top of conventional cryptographic countermeasures. This makes \ac{ptp}sec fully resilient against all known attacks against \ac{ptp}.
The experimental results show that our approach successfully identifies fine-grained asymmetry changes with microsecond accuracy and minimal detection time in all evaluated scenarios.
%
The authors have provided public access to their code and/or data at \url{https://github.com/tum-esi/ptpsec}.

\iftrue
\section*{Acknowledgment}
This work has received funding from The Bavarian State Ministry for the Economy, Media, Energy and Technology,
within the R\&D program „Information and Communication Technology”, managed by VDI/VDE Innovation + Technik GmbH.
Also, this work is supported by the European Union-funded project CyberSecDome (Agreement No.: 101120779).
\fi

\begin{spacing}{1.05}
\setlength\bibitemsep{3.5pt}

\bibliographystyle{IEEEtran}
\bibliography{reference}
\end{spacing}

\end{document}